\pdfoutput=1
\documentclass[twocolumn]{revtex4}
\usepackage{latexsym}
\usepackage{amsfonts}
\usepackage[latin1]{inputenc}
\usepackage[caption=false]{subfig}
\usepackage{graphicx}
\usepackage{mathrsfs}
\usepackage{amssymb}
\usepackage{amsmath}
\usepackage{color}
\usepackage{fancyhdr}
\newcommand{\gtap}{\stackrel{\displaystyle >}{\,_{\! \,_{\displaystyle
\sim}}}}  

\usepackage{color}

\begin{document}

\title{LHC physics of extra gauge bosons in the 4D Composite Higgs Model}

%

\author{D. Barducci, A. Belyaev, S. Moretti}
\affiliation{School of Physics and Astronomy, University of Southampton, Highfield, SO17 1BJ, UK.}
\email{d.barducci,a.belyaev,s.moretti@soton.ac.uk}
\author{S. De Curtis}
\affiliation{INFN, Sezione di Firenze,
Via G. Sansone 1, 50019 Sesto Fiorentino, Italy.}
\email{decurtis@fi.infn.it}
\author{G.M. Pruna}
\affiliation{Paul Scherrer Institute, CH-5232 Villigen PSI, Switzerland.}
\email{giovanni-marco.pruna@psi.ch}

\begin{abstract}
We study the phenomenology of both the Neutral Current (NC) and Charged Current (CC)
Drell-Yan (DY) processes at the Large Hadron Collider (LHC) within a
4 Dimensional realization of a Composite Higgs model with partial compositness
by estimating the integrated and differential event rates and taking
into account the possible impact of the extra fermions present in the spectrum.
We show that, in certain regions of the parameters space, the multiple neutral resonances present in the model
can be distinguishable and experimentally accessible in the invariant or transverse mass distributions.
\end{abstract}

\maketitle

\thispagestyle{fancy}


\section{Introduction}
\label{intro}
The Drell-Yan mechanism is one of the most important probes in the search for 
new vector boson resonances associated to possible physics Beyond the Standard Model
(BSM) and the LHC offers a unique chance to test DY phenomenology in high energy proton proton scattering allowing
to test BSM models with extra gauge bosons.

In this proceeding we focus on the analysis of the extra gauge bosons present in a particular BSM scenario with a Composite
Higgs state.

Composite Higgs models, where the physical Higgs state arises as a pseudo Nambu-Goldstone boson,
provide an elegant solution to the hierarchy
problem present in the Standard Model (SM) and suggest an alternative pattern leading to the mechanism of spontaneous
electroweak symmetry breaking. Moreover, in this scenario the Higgs potential is generically
a computable quantity and the Higgs
mass is no longer a free parameter.

The general structure is based on an idea proposed in the '80 \cite{Kaplan:1983fs} and was then specialized on the minimal
coset $SO(5)/SO(4)$ in \cite{Agashe:2004rs}.
A 4-Dimensional (4D) realization of this BSM model, called the 4D Composite Higgs Model (4DCHM), was presented in in \cite{DeCurtis:2011yx}
and is the framework within which we base our analysis.

The 4DCHM spectrum consists, beside the SM content, of nine extra gauge bosons: 5 neutral (denoted as $Z^\prime$) and 3 charged (denoted as $W^\prime$) with masses and couplings
described by two parameters, the compositeness scale $f\gg v \simeq 246$ GeV and the coupling constant $g_{\rho}$
of the extra gauge fields, $4 \pi \gg g_{\rho} \gg g$, where $g$ is the SM gauge coupling.
The spectrum also present extra fermions with both SM, herein collectively called
$t^\prime,b^\prime$, and exotic electric charges (5/3 and -4/3).

\section{Drell-Yan analysis}
\label{sec-1}
The DY NC $p~p\rightarrow l^+~l^-$ and CC $p~p \rightarrow l\nu_l+c.c.$
processes for the 4DCHM  have been studied in the context of the 14 TeV stage of the LHC in \cite{Barducci:2012kk,Barducci:2012as,Barducci:2013vva}.

Due to the large particle spectrum the model has been implemented in automatic tools such as
LanHEP \cite{Semenov:2010qt}, a package for the automatic generation of Feynman rules,
and CalcHEP \cite{Belyaev:2012qa}, a package for the automatic calculation of physical observables (cross sections, widths...)
to perform a fast phenomenological analysis.

Constraints on the parameter space due to electroweak precision tests have been taken into account
by requiring $m_{\rho} = m_{Z^\prime,W^\prime} \simeq f~g_{\rho}\gtap2\text{ TeV}$ and other physical constraints
such as top, bottom and Higgs mass have been implemented during the parameter scan.

\begin{figure}[h!]
\includegraphics[width=6.5cm]{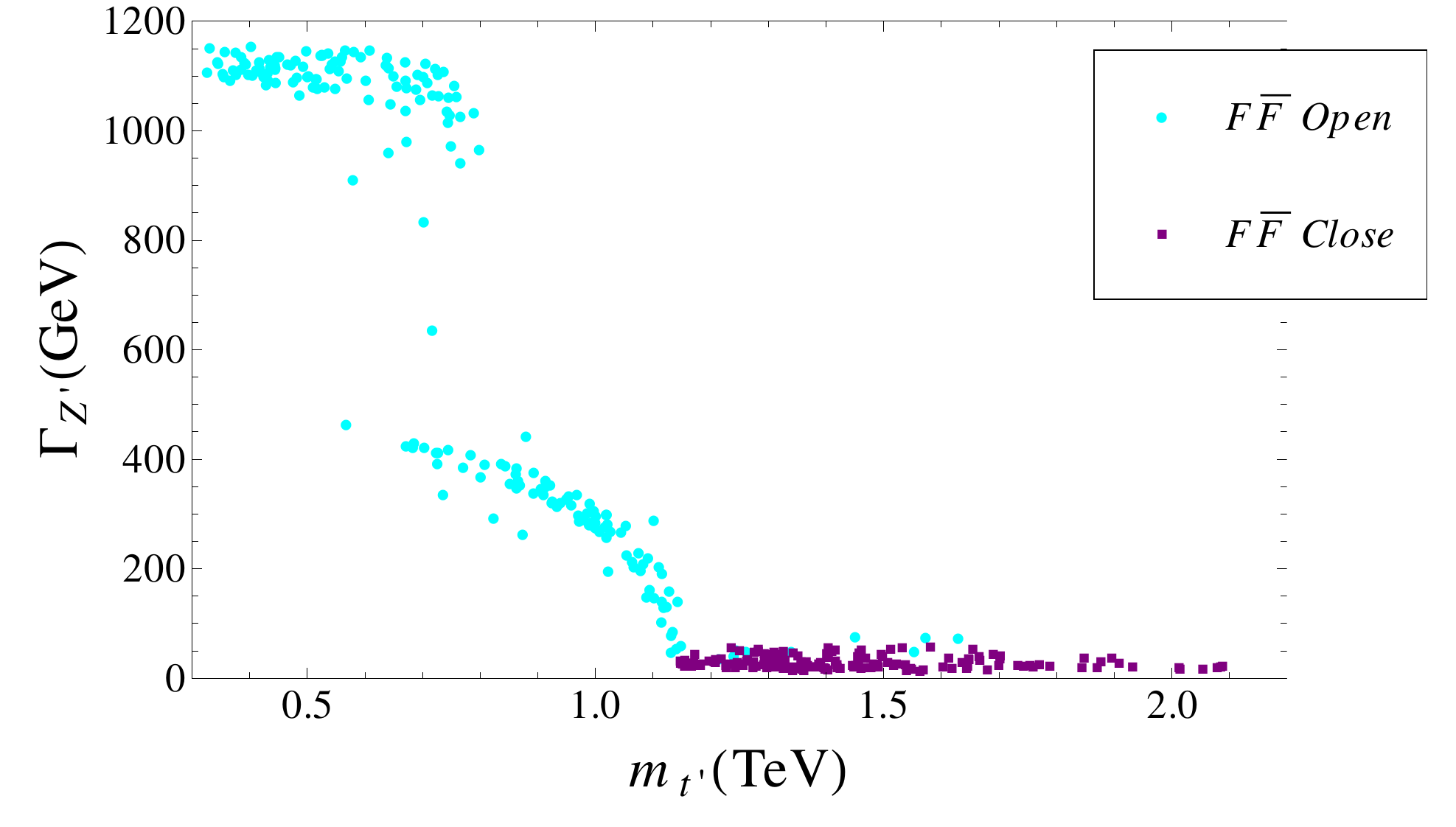}
\includegraphics[width=6.5cm]{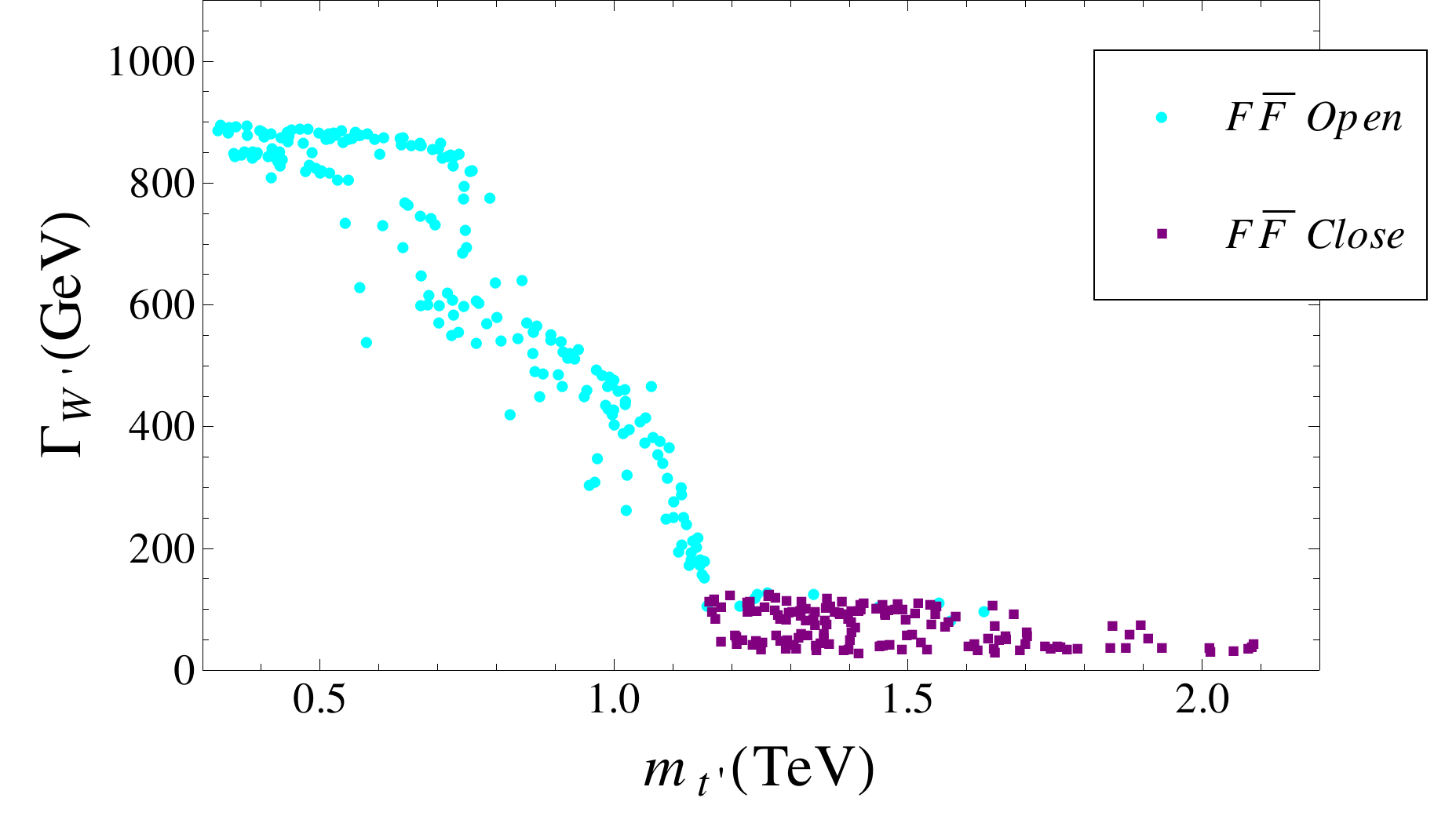}
\caption{Width of the lightest $Z^\prime$ and $W^\prime$ as a function of the mass of the lightest $t^\prime$. In purple points
where the decay in a pair of heavy fermions $F\bar F$ is kinematically not allowed, in cyan the points where this process is allowed. }
\label{fig:width}
\end{figure}

Despite not directly involved in the DY process the presence of the extra fermions is of great importance in our analysis.
In fact the mass of the lightest 4DCHM quark partner can dramatically affect the width of the extra gauge bosons, as shown in
Fig. \ref{fig:width}, due to the fact that, if kinematically allowed, the decay of the extra gauge bosons into the new fermionic
resonances gives a large contribution to the total width of the extra gauge bosons.

Taking in account this fact it is clear that the DY analysis could be used to test the gauge sector
of the 4DCHM in regions of the parameter space where these decay channels are not allowed, that is, where the
mass of the lightest extra fermions is at least half of the mass of the extra gauge bosons and so the widths
of the latter are small compared to their mass.

\begin{figure}[h!]
 \begin{center}
\includegraphics[width=4.7cm]{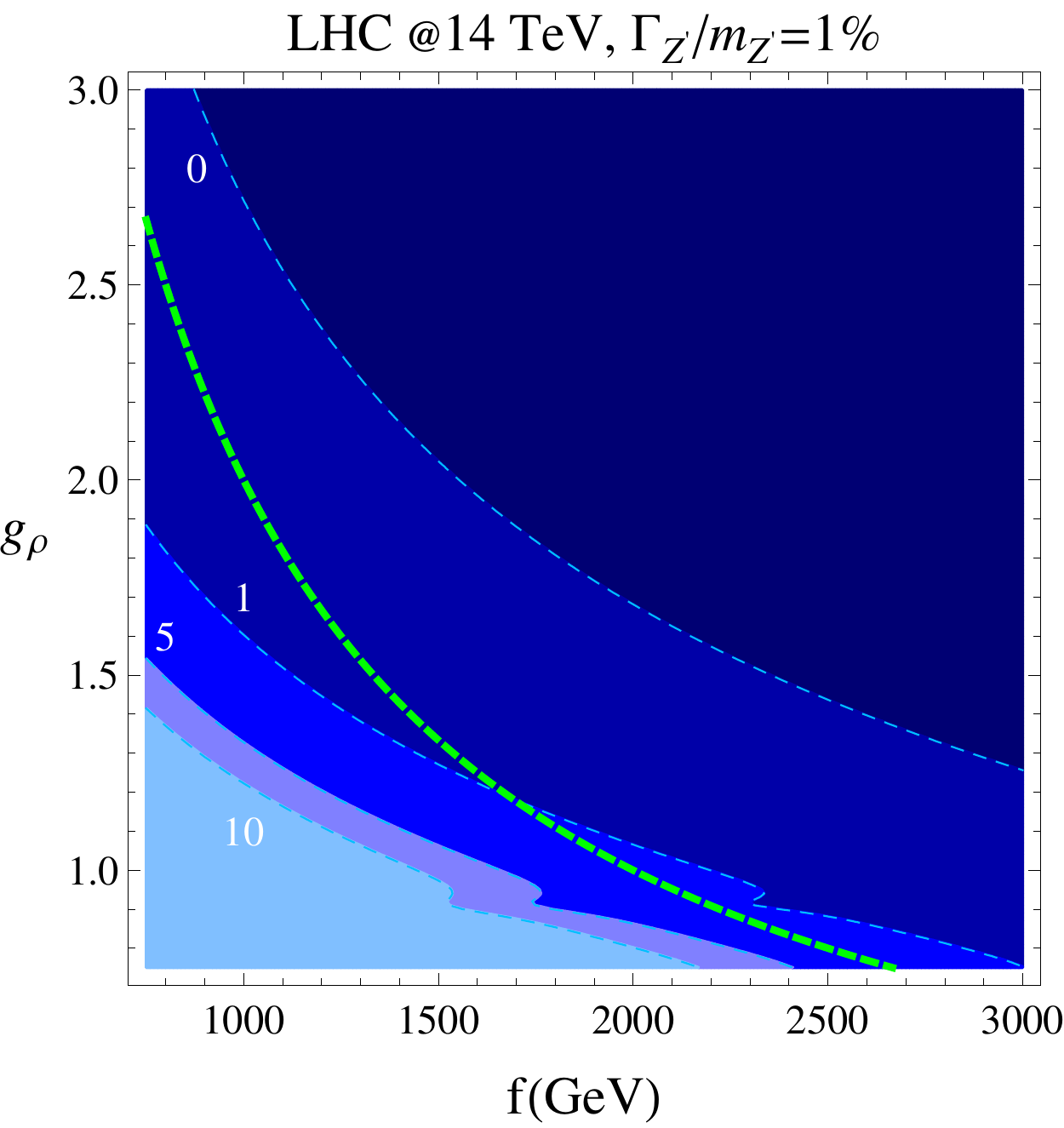}
\includegraphics[width=4.7cm]{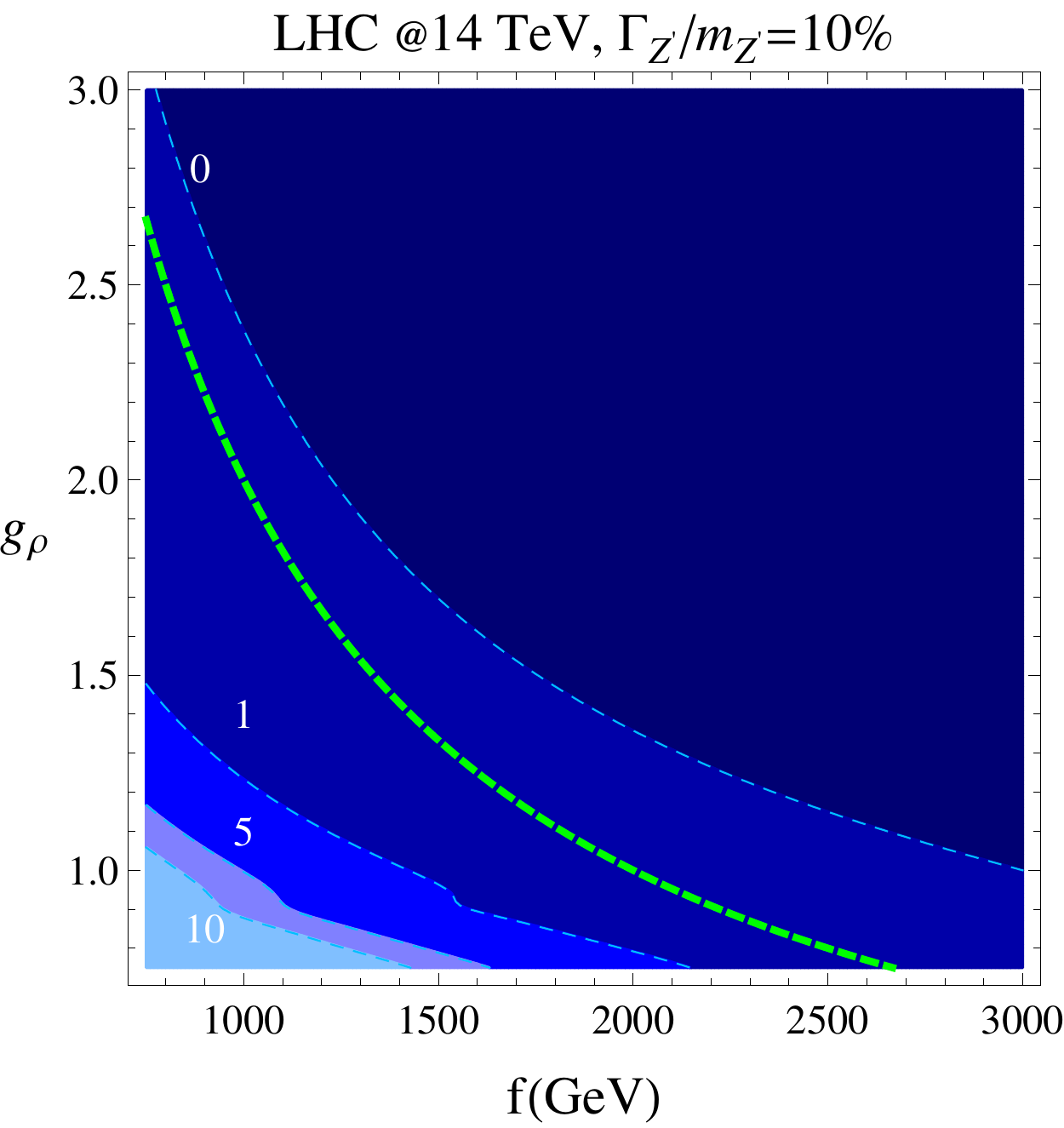}
\caption{Contour plot of $S/\sqrt{B}$ $[fb^{1/2}]$
in the $f,g_{\rho}$ plane for two different values of the extra gauge bosons width,
with respect to their masses: 1\% (top) and 10\% (bottom).
The green dashed lines represent the level curve where $f~g_{\rho}\simeq m_{Z^\prime,W^\prime}=2~\rm TeV$.}
\label{fig:sign}
\end{center}
\end{figure}

This is particularly relevant for the NC mode, for which the resonant mass can be reconstructed. Hence,
in Fig. \ref{fig:sign}
we show the contour plot in the plane $f-g_{\rho}$ for the quantity $S/\sqrt{B}$ where $B$ is the standard model background
and $S$ is the 4DCHM signal for two realistic choices of the $Z^\prime$ ratio $\Gamma_{Z'}/M_{Z'}$: 1\% and 10\%.
To obtain the statistical significance is enough to multiply the contour value for $\sqrt{L\varepsilon}$ where $L$
is the luminosity in $fb^{-1}$ and $\varepsilon$ is the efficiency to tag the final state.

\begin{figure}[h!]
\begin{center}
\includegraphics[width=4.2cm,angle=90]{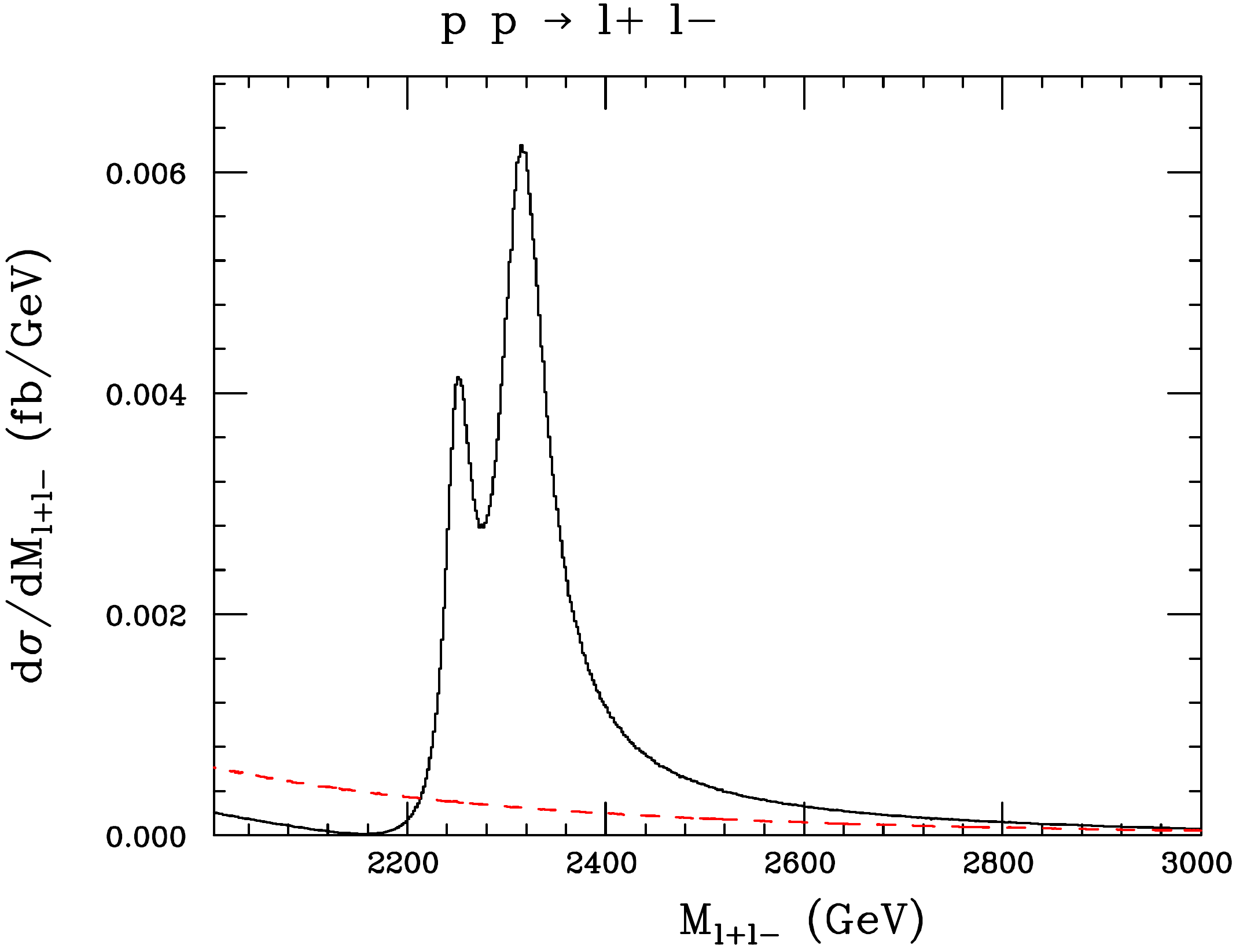}
\includegraphics[width=4.2cm,angle=90]{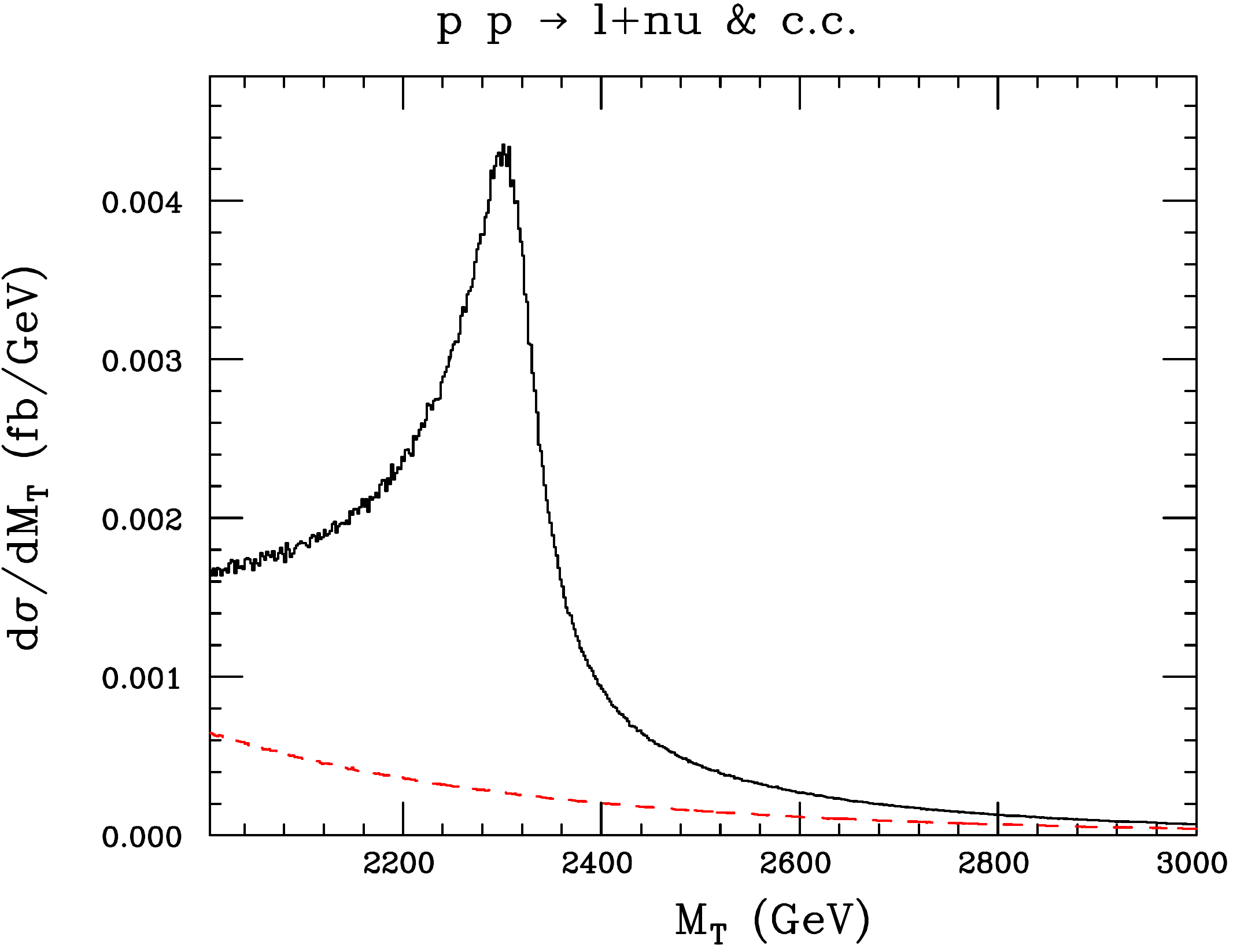}
\includegraphics[width=4.2cm,angle=90]{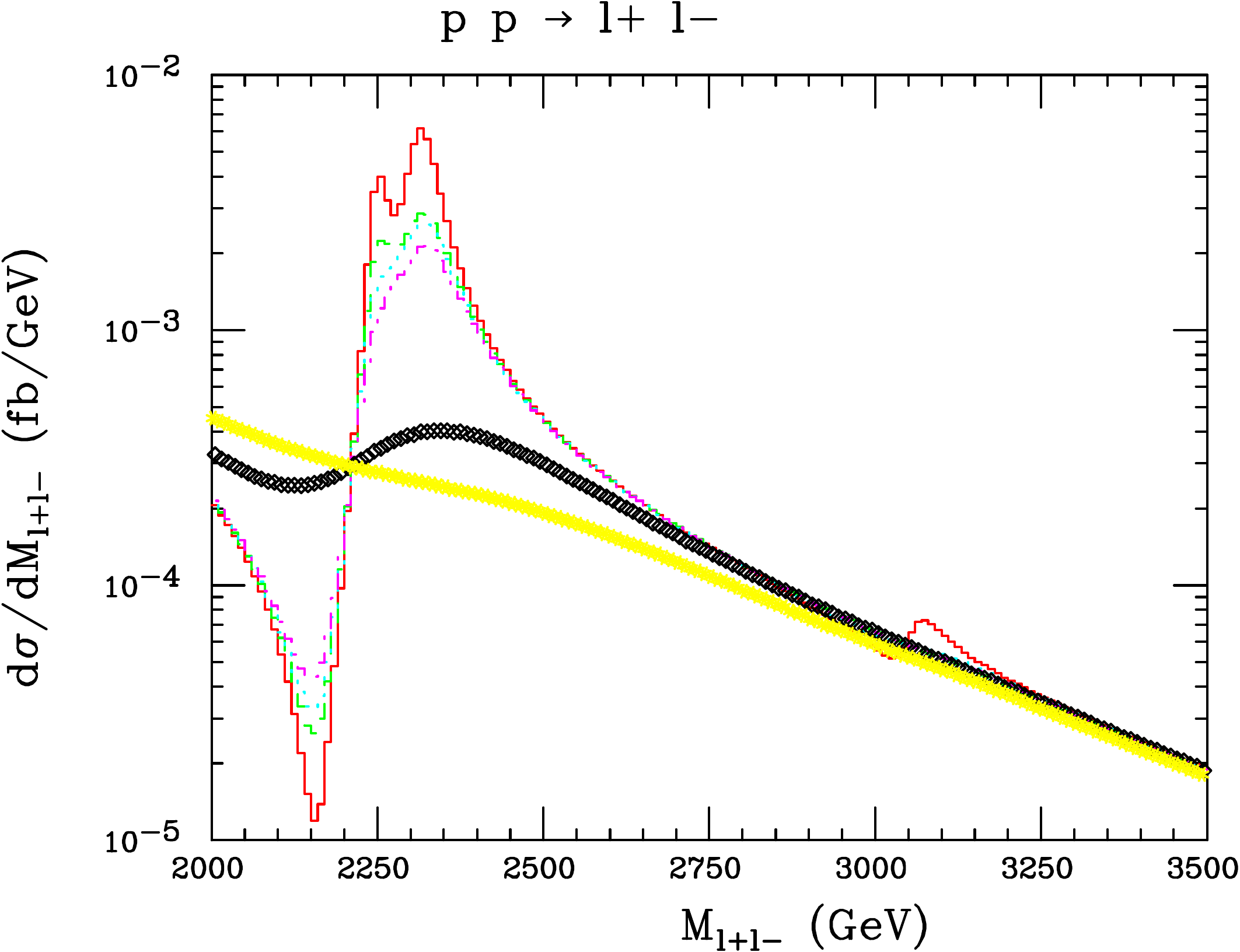}
\caption{Differential cross section in invariant mass for the NC (top) and in transverse mass for the
CC (middle) channels for $f=1.2$ TeV and $g_{\rho}=1.8$.
Integrated cross sections for the 4DCHM[SM], given as solid/black[dashed/red] lines, are 0.78[0.21] fb (NC)
and 1.11[0.23] fb (CC).
Line shape for the NC process (bottom) for the same combination of $f$ and $g_{\rho}$
and for different values of the masses of the lightest $t^\prime$ (see Ref. \cite{Barducci:2012kk} for the meaning
of colors in the last plot). Results are for the 14 TeV LHC setup.}
\label{fig:crosssections}
\end{center}
\end{figure}

Following the previous discussion we show in Fig. \ref{fig:crosssections} the differential distribution
in invariant (a) and transverse (b) mass for the NC and CC process, respectively,
for the choice $f=1.2~$TeV and $g_{\rho}$=1.8 with extra fermions heavy enough
such that no decay channel in a pair of the latter is open, while in (c) we show for the NC case only how the
the presence of lighter heavy fermionic resonances affects the line shape of the reconstructed $Z'$ mass distribution until the signal
does not emerge anymore over the background.

\section{Conclusion}
\label{sec-2}

In this proceeding we have shown how the 14 TeV stage of the LHC places us in the position of studying the rich
phenomenology of the gauge sector of the 4DCHM thereby making the latter testable for masses of the extra gauge bosons
up to 2-3 TeV.
Moreover, the two lighter neutral resonances are accessible in certain regions of the parameter space where the masses
of the extra fermions are not too light and so that the widths of the $Z^\prime$ are small compared to their masses.

Conversely the possible presence of extra fermions with a mass lower than the TeV scale represents the first possibility
to test the 4DCHM already with the 7 and 8 TeV LHC data.

The results are obtained for the 14 TeV LHC but  remain essentially valid for the next LHC run at 13 TeV.

\section*{Acknowledgments}
DB, AB and SM thank the NExT Institute for partial support.

\newpage
%
 \bibliography{proceeding_a}
\end{document}